# Reducing the Makespan in Hierarchical Reliable Multicast Tree[1]

Sang-Seon Byun,

Address: Department of Computer Science, Korea University, An-am dong 5th street, Seong-buk, Seoul, Korea

Email: ssbyun@os.korea.ac.kr

Telephone: +82-2-3290-3639

Fax: +82-2-922-6341

Chuck Yoo

Address: Department of Computer Science, Korea University, An-am dong 5th street, Seong-buk, Seoul, Korea

Email: hxy@os.korea.ac.kr

Telephone: +82-2-3290-3639

Fax: +82-2-922-6341

[1] This research was supported by the MIC(Ministry of Information and Communication), Korea, under the ITRC(Information Technology Research Center) support program supervised by the IITA(Institute of Information Technology Assessment)

**Abstract**

In hierarchical reliable multicast environment, makespan is the time that is required to fully and successfully transmit a packet from the sender to all receivers. Low makespan is vital for achieving high throughput with a TCP-like window based sending scheme. In hierarchical reliable multicast methods, the number of repair servers and their locations influence the makespan. In this paper we propose a new method to decide the locations of repair servers that can reduce the makespan in hierarchical reliable multicast networks. Our method has a formulation based on mixed integer programming to analyze the makespan minimization problem. A notable aspect of the formulation is that heterogeneous links and packet losses are taken into account in the formulation. Three different heuristics are presented to find the locations of repair servers in reasonable time in the formulation. Through simulations, three heuristics are carefully analyzed and compared on networks with different sizes. We also evaluate our proposals on PGM (Pragmatic General Multicast) reliable multicast protocol using ns-2 simulation. The results show that the our best heuristic is close to the lower bound by a factor of 2.3 in terms of makespan and by a factor of 5.5 in terms of the number of repair servers.

## I. INTRODUCTION

Recently many reliable multicast protocols deploy hierarchical structure to achieve scalability. Hierarchical reliable multicast (HRM) constitutes and maintains a control tree that is used to transmit recovery and feedback (ACK/NACK) apart from the multicast delivery tree. On the control tree, there must be repair servers that play a role of local recovery and feedback consolidation. Repair server can be set up as an exclusive server [1-3, 21, 22] or designated among adequate receivers [4-5, 23].

The performance of HRM can be evaluated by the bandwidth overhead due to local recovery and feedback consolidation and the makespan [6]. Makespan in HRM is defined as the time that is required to fully and successfully transmit a packet from a sender to all receivers. The bandwidth overhead and makespan are affected by the locations and the number of repair servers. For example, if a repair server is adjacent with a receiver that is susceptible to packet losses and the repair server experiences fewer packet losses than the receiver, the recovery and feedback traffics are limited to the server's domain. Additionally, if the delivery delays between the sender and each receiver are heterogeneous and additional delays are required to recover some packet losses, the arrival time of the packet at each receiver may be different. Therefore appropriate locating repair servers can reduce packet recovery time, and thus better makespan is obtained.

Low makespan is essential for achieving high throughput in window-based multicast sending schemes. In window-based schemes that provide full reliability, advancing the window is highly dependent on the slowest receiver's successful reception [7-9]. The slowest receiver is the member that takes the longest time to receive a packet from the sender, and this time becomes equivalent with the makespan. For our best knowledge, there is no previous work to improve the makespan in HRM networks.

In this paper we propose a new method that can extract locations of repair servers to reduce makespan in heterogeneous HRM networks. First we formulate this problem as an MIP (mixed integer programming). Since it is impossible to obtain optimal solution of our MIP formulation, we propose several heuristics that can provide good solutions to this problem in reasonable time. With our system, Pentium IV 3.0GHz and 1.5GB Memory, it took about 4 hours GNU glpk [14] to locate only 5 repair servers on network of 50 nodes. In case of 100 nodes, glpk could not solve integer solution until 24 hours have elapsed. If the network size is n and k proxies are to be located, our MIP has the time complexity of $O((n-k)^n)$ in the worst case. Thus the computation time increases exponentially as network size grows. The simulation results show that the greedy heuristic of locating a repair server first on the node that lies on the longest path and has the incident edge showing the largest expected delivery delay is best among our heuristics. In addition the best heuristic is close to the lower bound obtained using LP relaxation by a factor of 2.3 in terms of makespan.

Our proposal can be adapted to the real application as follows:

For instance, a service provider may perform file distribution or multi-media streaming using IP multicast or overlay multicast. The service provider collects the statistics of each link – per-link delay and per-link loss rate – and defines appropriate makespan in accordance with the application type and user's requirement. Then using our proposal, they can decide the feasible locations of the repair proxies in order to achieve the makespan in reasonable time. In addition, due to the fluctuation of link statistics and the dynamic membership, the optimal locations of repair proxies must be recomputed as fast as possible.

This paper is organized as follows. In Section 2 we present the impact of makespan in the existing window-based sending schemes. In Section 3 we present the HRM model used in this paper and describe an expected delivery delay model that reflects the locations of the repair servers and heterogeneity of network environment. In Section 4 we provide the MIP that formulates our model. In Section 5 we propose three different heuristics that can provide solutions in reasonable time, and the MIP for computing the lower bound in terms of the number of repair server. Section 6 provides simulation results on the heterogeneous networks with various sizes and the result of simulation experiment on PGM (Pragmatic General Multicast) [12]. Finally, concluding remarks are given in Section 7.

## II. WINDOW-BASED SENDING SCHEMEWindow-based Sending Schemes

PGMCC (Pragmatic General Multicast Congestion Control) [7] is a TCP-like window-based congestion control scheme that operates on an HRM protocol, PGM (Pragmatic General Multicast). PGM deploys routers that perform local recovery and feedback consolidation. PGMCC elects a group representative (acker) and mimics the TCP congestion window mechanism between the sender and the acker. The acker is chosen as the receiver with the worst throughput among group members. The acker election process is performed by measuring packet loss rate and RTT (Round Trip Time). RTT is measured through computing the difference between the most recent sequence number sent and the highest known sequence number to a receiver that is noted in feedback packet. Therefore, if the makespan is reduced through adequate placement of PGM-enabled routers, the RTT between the sender and the acker can be reduced also, and thus the congestion window can advance faster.

MTCP (Multicast TCP) [8] is another congestion control scheme using hierarchical structure for large-scale reliable multicast. In MTCP, SA(Sender Agent)s are deployed to perform local recovery and feedback consolidation for their children, and between SAs, and between SA and leaf nodes, TCP-like congestion control is performed. The congestion window in each SA is incremented when each SA receives ACKs for a packet from all of its children. Therefore the reduced makespan can be beneficial to the fast advance of the congestion window. In RMTP-II (Reliable Multicast Transport Protocol-II) [9], when receivers and DR (Designated Receiver)s receive packet successfully, they respond by generating ACKs and propagating them up the control tree, and back to the sender. DR also performs ACK consolidations and local recovery. When the sender gets an ACK for a data packet it has sent, it can advance its transmission window and free that packet from memory. Therefore, if the makespan becomes shorter, the transmission window can advance faster, and this results in higher throughput and lower memory usage.

In order to alleviate throughput degradation due to the worst receiver, multi-rate sending schemes can be deployed. Chaintreau [16] analytically proves that window-based sending scheme might be harmful with reliable multicast transmission. In addition they show that this result extends to unreliable application that may find it difficult to manage a 50% drop in the average throughput when the number of receiver increases, and multi-rate sending schemes are best suited to multicast sessions with a large number of receivers. However, the flexibility of multi-rate scheme is paid in terms of coding costs, some bandwidth inefficiency, and possibly a coarser match of sender and receiver data rate [7].

## III. HRM AND EXPECTED DELIVERY DELAY MODEL

In this section, we describe HRM and expected delivery delay model. As described in section 1, delivery delay is the time required to successfully transmit a packet from the sender to a receiver. In order to extract locations of repair servers to reduce makespan, HRM and expected delivery model must be established reflecting heterogeneity and locations of repair servers.

### A. HRM Model
This paper assumes the HRM model of the following characteristics (Figure 1).

(1) If a repair server is placed in a node, then the node itself becomes a repair server.
(2) Consider a single-sender multicast tree where the root is the unique sender, all leaves are receivers and all intermediate nodes can be repair server [1, 2, 3].

(3) The topology of control tree is identical to that of its underlying multicast tree (IP multicast tree), and loss probabilities and delays at the links of the control tree are given. It is indeed possible to acquire knowledge about the underlying multicast tree even under dynamic membership situation. Levine et al. [10-12, 19] describe a way of establishing a control tree that is identical to its underlying multicast tree and collecting link loss statistics. Especially, Caceres et al. [17-18] propose methods to infer per-link loss rates and delays. Even if topology and loss rates are not known in real world, our method may still be useful for comparison and assessment purpose.

(4) The control tree is partitioned into subtrees that form a hierarchy rooted at the sender. All nodes in a subtree are combined into a subgroup, and each subgroup has a repair server located at its root. The sender itself is a repair server by default. These features are deployed in [1-3] [5] [7] [13] [23].

(5) Repair server multicasts the original data to its own subgroup. Each receiver sends feedback (NACK or intermediate ACK) to its own repair server when a packet loss is detected, and the repair server retransmits the lost packet to the whole subgroup. We assume that all feedback packets are delivered via an out-of-band channel, so all feedback packets are delivered safely to repair servers. This assumption is also used in [20].

(6) Feedbacks and transmissions/retransmissions are limited only between a repair server and the receivers of its subgroup and they do not reach receivers/repair servers of any other subgroup. For this purpose, a new multicast address per subgroup is assigned [2], or TTL (Time to live) may be used to scope subgroup [4]. Additionally subcasting and TTL scoping can be used simultaneously [5].

*B. Expected Delivery Delay Model*

We explore an expected delivery delay model that reflects heterogeneity and locations of repair servers. We assume that per-link delay and loss rate are measured in end-to-end manner. There are some previous work [17-18] that describes the methods to infer per-link loss rate and delay in end-to-end manner. If we assume that the per-link delay is inferred using these methods, the inferred delay must include both queuing and propagation delay. Additionally, as in real network, if we assume that each router and node processes not only the multicast packet but also its own background traffic, the queuing delay and packet drop rate of each router and node distributes almost randomly. And also, we emphasize the successful arrival of a packet, thus parameter $t$ of equation (1) denotes a current delay incurred from congestion/flow control. Also it is assumed that packet losses at each link are independent.

Figure 2 shows expected delivery model that reflects heterogeneity of per-link delay and loss rate. In Figure 2, <p, d> on each link represents <per-link loss rate, per-link delay>.

A summary of the used notations is given in Table 1.

$E(N_{a,b})$ can be obtained as follows, where $t$ is the reciprocal of sending rate:

$$E(N_{a,b}) = \frac{1 - \prod_{i \in \psi(a,b)}(1-p_i)}{\prod_{i \in \psi(a,b)}(1-p_i)} \times t + L_{a,b} \tag{1}$$

$E(D_{S, R(w)})$ can be expressed in the following two cases.

*1) in case* $\pi(S, R(w)) = \{S\}$ *(S is unique repair server between S and R(w)):*

$$E(D_{S, R(w)}) = \left[\prod_{i \in \psi(S, R(w))} (1-p_i)\right] L_{S, R(w)} + \left[1 - \prod_{i \in \psi(S, R(w))} (1-p_i)\right] \left(E(N_{S, R(w)}) + E(D_{S, R(w)})\right) \quad (2)$$

By eliminating $E(D_{S, R(w)})$ at the right side of equation (2), we obtain

$$E(D_{S, R(w)}) = L_{S, R(w)} + \frac{1 - \prod_{i \in \psi(S, R(w))} (1-p_i)}{\prod_{i \in \psi(S, R(w))} (1-p_i)} E(N_{S, R(w)}) \quad (3)$$

*2) in case* $\theta(S, R(w)) \geq 2$ *and* $node(j) \in \pi(S, R(w))$

$$E(D_{S, R(w)}) = \left[\prod_{i \in \psi(S, node(j))} (1-p_i)\right] \left(L_{S, node(j)} + E(D_{node(j), R(w)})\right)$$
$$+ \left[1 - \prod_{i \in \psi(S, node(j))} (1-p_i)\right] \left(E(N_{S, node(j)}) + E(D_{S, R(w)})\right) \quad (4)$$

$$E(D_{S, R(w)}) = L_{S, node(j)} + \frac{1 - \prod_{i \in \psi(S, node(j))} (1-p_i)}{\prod_{i \in \psi(S, node(j))} (1-p_i)} E(N_{S, node(j)}) + E(D_{node(j), R(w)}) \quad (5)$$

By definition of $E(D_{S, R(w)})$,

$$L_{S, node(j)} + \frac{1 - \prod_{i \in \psi(S, node(j))} (1-p_i)}{\prod_{i \in \psi(S, node(j))} (1-p_i)} E(N_{S, node(j)}) \equiv E(D_{S, node(j)}) \quad (6)$$

so $E(D_{S, R(w)})$ can be written as

$$E(D_{S, R(w)}) = E(D_{S, node(j)}) + E(D_{node(j), R(w)}) \quad (7)$$

And if $node(k) \in \pi(node(j), R(w))$, we obtain

$$E(D_{node(j), R(w)}) = E(D_{node(j), node(k)}) + E(D_{node(k), R(w)}) \quad (8)$$

Thus by setting, $\pi(S, R) = \{S, proxy_0, proxy_1, ..., proxy_z\}$, we obtain an recursive form as follows:

$$E(D_{S, R}) = E(D_{S, proxy_0}) + E(D_{proxy_0, R})$$
$$E(D_{proxy_0, R}) = E(D_{proxy_0, proxy_1}) + E(D_{proxy_1, R})$$
$$.$$
$$.$$
$$.$$
$$E(D_{proxy_{z-1}, R}) = E(D_{proxy_{z-1}, proxy_z}) + E(D_{proxy_z, R}) \quad (9)$$

So using this recursive form, we compute expected delivery delays of all receivers if set of repair servers is configured. Therefore this expected delivery delay model is used to compute the makespan.

## IV. MIXED INTEGER PROGRAMMING FORMULATION

The problem of minimizing makespan in HRM can be formulated as follows:

Given a tree $T(V, E)$, let $n=|V|$ and $V_L$ be the set of leaf nodes in $T$, and let $m=|V_L|$. Find optimal locations of $k$-repair servers in $T$ to minimize the makespan.

The above formulation can be cast into an MIP formulation. The notations for the MIP are given as follows:

$x_{ij}$ : $v_j$ is a repair server of $v_i$, and $v_i$ must be either a repair server or leaf node.

$y_j$ : $v_j$ is selected as a repair server.

$z_i$ : $v_i$ is a leaf node.

$r_{ij}^k$ : Both $v_i$ and $v_j$ are on the path to the $v_k$ from sender, and $v_i$ is an ancestor of $v_j$.

$p_{ij}$ : Both $v_i$ and $v_j$ are on the same path.

$w_{ij}^l$ : $v_l$ lies on the path between $v_i$ and $v_j$, and $v_j$ is an ancestor of $v_i$.

$d_{ij}$ : expected delivery delay between $v_i$ and $v_j$, and calculated using the expected delivery delay model in sub-section 3.2

Our objective is:

$$\text{Minimize } Makespan$$

subject to

$$\sum_{j=0}^{n-1} y_j = k \quad \text{Constraint (1)}$$

$$x_{ij} \leq y_j \quad \text{for each } v_i, v_j \in V \quad \text{Constraint (2)}$$

$$x_{ij} \leq y_i + z_i \quad \text{for each } v_i, v_j \in V \quad \text{Constraint (3)}$$

$$x_{ij} \leq 1 - w_{ij}^l y_l \quad \text{for each } v_i, v_j, v_l \in V \quad \text{Constraint (4)}$$

$$x_{ij} \leq p_{ij} \quad \text{for each } v_i, v_j \in V \quad \text{Constraint (5)}$$

$$\sum_{j=0}^{n-1} \sum_{i=0}^{n-1} x_{ij} = k + m - 1 \quad \text{Constraint (6)}$$

$$\text{Makespan} \geq \sum_{j=0}^{n-1}\sum_{i=0}^{n-1} d_{ij} x_{ij} r_{ij}^k \quad \text{for each } v_k \in V_L \quad \text{Constraint (7)}$$

$$x_{ij} \in \{0,1\} \text{ for each } v_i, v_j \in V \quad \text{Constraint (8)}$$

$$y_j \in \{0,1\} \text{ for each } v_j \in V \quad \text{Constraint (9)}$$

$$z_i \in \{0,1\} \text{ for each } v_i \in V \quad \text{Constraint (10)}$$

$$w_{ij}^l \in \{0,1\} \text{ for each } v_i, v_j, v_l \in V \quad \text{Constraint (11)}$$

$$p_{ij} \in \{0,1\} \text{ for each } v_i, v_j \in V \quad \text{Constraint (12)}$$

$$r_{ij}^k \in \{0,1\} \text{ for each } v_i, v_j \in V \text{ and for each } v_k \in V_L \quad \text{Constraint (13)}$$

The constraint (1) makes sure that the number of available repair servers is limited to $k$. The constraint (2) guarantees that a node $v_j$ must be a repair server if some other node $v_i$ receives repair packets from the node $v_j$. The constraint (3) ensures that a node $v_i$ must be a repair server or leaf node if the binary variable $x_{ij}$ is 1. Also, through the third constraint, the constraint (7) guarantees that the expected delivery delay is computed on the path only between repair servers or between a repair server and a leaf node. The constraint (4) makes sure that no repair server exists on the path between a node $v_j$ and a node $v_i$ except the node $v_j$ and $v_i$ if the node $v_i$ receives retransmissions form the node $v_j$, and also guarantees that the node $v_j$ is an ancestor of the node $v_i$. The constraint (5) ensures that a node $v_i$ is assigned to a node $v_j$ that lies on the same path with the node $v_i$. The constraint (6) assures that the number of sections on that the expected delivery delays are computed is same with the number of sections between repair servers or a repair server and a leaf node. Finally the constraint (7) ensures that the maximum expected delivery delay cannot exceed the makespan.

## V. HEURISTICS FOR THE GREEDY SELECTION OF REPAIR SERVERS

We propose three heuristics in this section. We also give a random selection of repair servers which serves as a base line for the evaluation of our three heuristics. In each heuristic, we assume that the sender node itself is a repair server by default as in other HRM methods. Therefore each heuristic selects additional *k-1* repair servers if we intend to select *k* repair servers in total.

### A. Max Expected Link Delay First: HMaxDelay

Our first heuristic is to use an expected delivery delay of outbound incident edge. In this paper, the words

'downward' and 'outbound' indicate the direction from sender to leaf, and upward indicates the opposite direction. We refer to this heuristic as *HMaxDelay*. It picks a node to which the largest expected delivery delay edge is incident. The greedy heuristic is based on the intuition that a large expected delivery delay can be brought by high packet loss rate, so if we place repair servers at the nodes whose incident edges have larger expected delivery delay than others, more reduced retransmission delays are obtained and this leads to more reduced makespan. Figure 3 illustrates the algorithm of *HMaxDelay* heuristic.

In a HRM tree $T(V, E)$, we assume that every node $v \in V$ is selected as repair server and then compute the expected delivery delay of each edge $e \in E$ using equation (9) described in subsection 3.2. In our model, expected delivery delay can be computed between repair servers, or repair server and leaf node. However, since we must obtain the expected delivery delay of each single edge, we just only assume that all nodes except leaf nodes are selected as repair servers. After that, select the node as repair server to which the highest expected delivery delay edge is incident. This process iterates till $k-1$ repair servers are found or no more repair server candidate is available.

For the *HMaxDelay* algorithm, if there are $n$ repair server candidates and we select $k-1$ repair servers among them, we must iterate the process of finding max delay edge $k-1$ times. Therefore it takes $\theta(k\log n)$. It takes $\log n$ steps to find the maximum key among $n$ elements.

### B. Max Out-degree First: HMaxDegree

Our second heuristic is to select repair servers based on node degree. We refer to this heuristic as *HMaxDegree*. It picks a node as repair server that has the highest degree. The intuition is that by selecting repair servers based on their own degree, more nodes can be serviced fast, and the number of repair servers required can be reduced. Therefore it is more probable to obtain reduced makespan. The algorithm of *HMaxDegree* heuristic is described in Figure 4. *HMaxDegree* takes $\theta(k\log n)$ steps also.

### C. Longest Path First: HLongPath

Let $P$ a set of paths from sender to all leaf nodes. The third heuristic is to find the longest path $p_i \in P$ - that has the largest expected delivery delay - and select a repair server on the path pi to which the largest expected delivery delay edge is incident. We note this heusristic as *HLongPath*. The intuition is that because makespan is determined by the longest path, we can reduce makespan through reducing the expected delivery delay of the longest path first. The *HLongPath* algorithm is illustrated in Figure 5.

First the algorithm finds the longest path $p_i$. Then we assume that every node in the path $p_i$ is a repair server and selects a repair server on the path $p_i$ whose outbound incident edge has the largest expected delivery delay. After selecting the repair server, if more repair servers are available, computing of the expected delivery delay of each path, finding the longest path and selecting repair server are repeated respectively until $k$ repair servers are – including the sender node – selected.

If there are $m$ leaf nodes, it takes $\log m$ step to find the longest path $p_i$. Then locating $j$ repair server candidates on the path $p_i$, takes *log j* step to find the node whose incident edge has the largest expected delivery delay. If all paths have j repair server candidates, and in every repetition, repair server is selected from the different path, this algorithm needs *(k-1) (log j + log m)* steps. Therefore this algorithm takes *O(klog m)* or *O(klog j)*.

## D. Lower Bound of Number of Repair Server

We would like to compare the required number of repair servers by our heuristic to the minimum number of repair servers assuming a given makespan. If the obtained makespan using our heuristic is applied to the following MIP formulation, we can compute the minimum required number of repair servers. Therefore we can evaluate our heuristic in terms of the required number of repair servers that satisfy the makespan.

The following MIP also cannot be solved in reasonable time due to its exponential time complexity so we obtain a crude lower bound (LP bound) and compare our heuristic with the lower bound. We can obtain the LP bound in a short time through LP-relaxation of the MIP. LP bound is obtained by removing all the integer constraint of the MIP. Thus, if the objective is minimization, the LP bound is the lower bound of the MIP integer solution.

$$\text{Minimize} \sum_{j=0}^{n-1} y_j$$

subject to

$$x_{ij} \leq y_j \quad \text{for each } v_i, v_j \in V \qquad \text{Constraint (1)}$$

$$x_{ij} \leq y_i + z_i \quad \text{for each } v_i, v_j \in V \qquad \text{Constraint (2)}$$

$$x_{ij} \leq 1 - w_{ij}^l y_l \quad \text{for each } v_i, v_j, v_l \in V \qquad \text{Constraint (3)}$$

$$x_{ij} \leq p_{ij} \quad \text{for each } v_i, v_j \in V \qquad \text{Constraint (4)}$$

$$\sum_{j=0}^{n-1} \sum_{i=0}^{n-1} x_{ij} = \sum_{j=0}^{n-1} y_j + m - 1 \qquad \text{Constraint (5)}$$

$$C \geq \sum_{j=0}^{n-1} \sum_{i=0}^{n-1} d_{ij} x_{ij} r_{ij}^k \quad \text{for each } v_k \in V_L \qquad \text{Constraint (6)}$$

$$x_{ij} \in \{0,1\} \text{ for each } v_i, v_j \in V \qquad \text{Constraint (7)}$$

$$y_j \in \{0,1\} \text{ for each } v_j \in V \qquad \text{Constraint (8)}$$

$$z_i \in \{0,1\} \text{ for each } v_i \in V \qquad \text{Constraint (9)}$$

$$w_{ij}^l \in \{0,1\} \text{ for each } v_i, v_j, v_l \in V \qquad \text{Constraint (10)}$$

$$p_{ij} \in \{0,1\} \text{ for each } v_i, v_j \in V \qquad \text{Constraint (11)}$$

$$r_{ij}^k \in \{0,1\} \text{ for each } v_i, v_j \in V \text{ and for each } v_k \in V_L \qquad \text{Constraint (12)}$$

## VI. SIMULATION RESULTS

In this section, we compare the performance of our heuristics on several different network sizes. Ideally we would like to observe how different our best heuristic is from the integer optimal solution. However we are unable to obtain the optimal solution even for a small network of just 50 nodes in reasonable time due to its exponential time complexity. Thus our best heuristic is compared with the lower bound obtained by LP relaxation.

We use ToGenD topology generator [15] for our simulation and generate topologies of 100, 200, 500, and 1000 nodes. In each topology, we define the node that has number 0 as a sender node and generate HRM tree using shortest path algorithm in order to minimize total delay of each sender-receiver pair. In each HRM tree, sender node is a repair server by default and every node can be selected as repair server except leaf nodes. Per-link delays are randomly assigned from the range of 10ms to 40ms and packet loss rates are chosen from the range of 0.0 to 0.1.

### A. Performance Comparison among Greedy Heuristics

Figure 6 shows measured makespan of each heuristic with respect to the number of repair servers on each network size. Each makespan is computed using the expected delivery delay model - equation (3) and (6) - described in Section 3.

As shown in Figure 6, we observe that the heuristic *HLongPath* outperforms other heuristics in terms of makespan regardless of network size. We can also see that if each makespan arrives at some lower limit, makespan does not decrease even though we increase the number of repair servers. For instance, if no more repair server candidates are available on the path that determines the makespan, additional placement of repair servers on other paths does not alter the makespan In case of network size of 100 nodes, available number of node that can be selected as a repair server is measured as 25. The heuristic *HRandom* arrives at the lower limit of makespan when 24 repair servers are selected. The *HMaxDelay*, *HMaxDegree* and *HLongPath* need 15, 22 and 10 repair servers respectively in order to arrive at the lower limit of makespan.

Table 2 shows the lower limit of makespan and the minimum number of repair servers to attain this lower limit with respect to each heuristic on each network size. The parenthesized number in the field of network size means available number of node that can be selected as a repair server in accordance with each network size. We observe that the *HLongPath* heuristic needs fewer repair servers than the other three heuristics in order to attain the lower limit of makespan.

### B. Comparison with Lower Bound

Since *HLongPath* performs the best among the heuristics, we only need to compare *HLongPath* with the lower bound from the LP relaxation. The LP-relaxed MIP is computed using the simplex method. Figure 7(a) plots the makespan of *HLongPath* and its lower bound. It is observed that the result of our best heuristic, *HLongPath* is close to the lower bound by a factor of 2.3 maximally in terms of the makespan. Figure 7(b) plots the minimum number of repair server when it is assumed that the makespans obtained by *HLongPath* are given. In our system, our GLPK cannot obtain the result of LP relaxation on the 200 nodes case. The lower bound obtained from LP relaxation of MIP in subsection 5.4 is plotted also. We see that our result is close to the lower bound by a factor of 5.5 in terms of the number of repair server when the given makespan is 113. 81ms. Based on the previous work with respect to MIP optimization [24], we regard the solution of a heuristic as relatively good, if the solution is close to the LP bound with the factor of less than about 5.

### C. Comparison with the Optimal Integer Solution in Small Size Network

We can obtain integer solution of our MIP using glpk in small size network. Thus we compare the result of our best heuristic with the integer solution in the network size of 25, 30, 40 and 50. The integer solution is obtained by the branch and bound. Table 3 shows the obtained results. First we compute the makespan of our heuristic, and obtain the proxy sizes that cause reduction of makespan. In network size of 25, 30 and 50, makespan reduction occurs at the proxy size of 3 and 4. In network size of 40, makespan is reduced at the proxy size of 3, 4 and 5. In network size of 25 and 30, there is no difference between our greedy heuristic and the optimal solution. However in network size of 40 and 50, the optimal solution shows better performance than our greedy method. In the network size of 40, our heuristic needs 5 proxies to obtain the makespan of 119.96, on the other hand the optimal solution needs 4 proxies. Thus, we can infer that as network size grows, the makespan difference between our heuristic and the optimal solution also increase. However, our heuristic can be useful to obtain relative good solution in reasonable time.

### D. Simulation Experiment on PGM

We simulate our proxy placement method to minimize makespan on PGM HRM tree using ns-2. Simulation setup process is described as follows: (1) create a simulation topology using ToGend and choose a sender node randomly. (2) Decide the optimal locations of proxies on the generated topology using our HRM delivery delay model and the best heuristic *HLongPath* to minimize the makespan. (3) Provide the role of PGM enabled router of PGM to the sender and proxies, and put on the PGM receiver agent to the receivers. Simulation parameters are same with those shown in table 4.

Figure 10 shows that the *HLongPath* heuristic has better makespan than other three heuristics at most of packets. *HLongPath* does not show best maekspan through entire packet sequence. However we find that *HLongPath* does not allow the makespan over 5 seconds through entire packet.

## VII. CONCLUSIONS

Many hierarchical reliable multicast methods deploy local recovery using repair servers to reduce the delay that is required to recover packet loss, and this delay is strongly affected by the locations of repair servers. We define makespan in HRM methods, as the time required to fully and successfully transmit a packet from the sender to all receivers. Makespan includes transmission delay and feedback delay, and if a packet loss is occurred, it includes recovery delay also. Thus makespan can be reduced through adequate placement of repair servers. Especially, in window-based sending schemes, the more makespan is reduced, the faster the transmission window advances, and therefore the better throughput is achieved.

In this paper, we propose a new method to place repair serves that can reduce the makespan in heterogeneous HRM networks. We use an expected delivery delay model to reflect heterogeneity and locations of repair servers, and formulate the makespan minimization problem into MIP. In addition we propose three heuristics except random placement policy in order to obtain the locations of repair servers in reasonable time. Through our simulations, we observe that the heuristic *HLongPath* is the best in that fewer repair servers are needed to attain the lower limit of makespan. For a network about 1000 nodes we can obtain the locations of repair servers in just a few scores of seconds, which will allow the network managers to identify the locations of repair servers in reasonable time when our method in this paper is deployed.

< List of figures and table captions >

**Fig. 1** HRM Model

**Fig. 2** Expected delivery delay model reflecting heterogeneity: Expected delivery delay between node *S* and *R(w)* when at each node *S*, 2, *j* and *k*, a repair server is located.

**Fig. 3** Heuristic Algorithm of *HMaxDelay*

**Fig. 4** Heuristic Algorithm of *HMaxDegree*

**Fig. 5** Heuristic Algorithm of *HLongPath*

**Fig. 6** Makespan of each heuristic as a function of the number of repair servers

**Fig. 7** Comparison of *HLongPath* in network size of 100 nodes

**Fig. 8** Comparison with optimal in network size of 40 and 50nodes

**Fig. 9** Makespan of each heuristics measured on PGM

**Table 1**. Notations for expected delivery model

**Table 2**. Each tuple - (**Lower limit of makespan (*ms*), minimum number of repair servers to attain this lower limit**) in accordance with each heuristic and network size.

**Table 4** Simulation parameters

**Table 3.** Makespan of our heuristic and the integer optimal solution

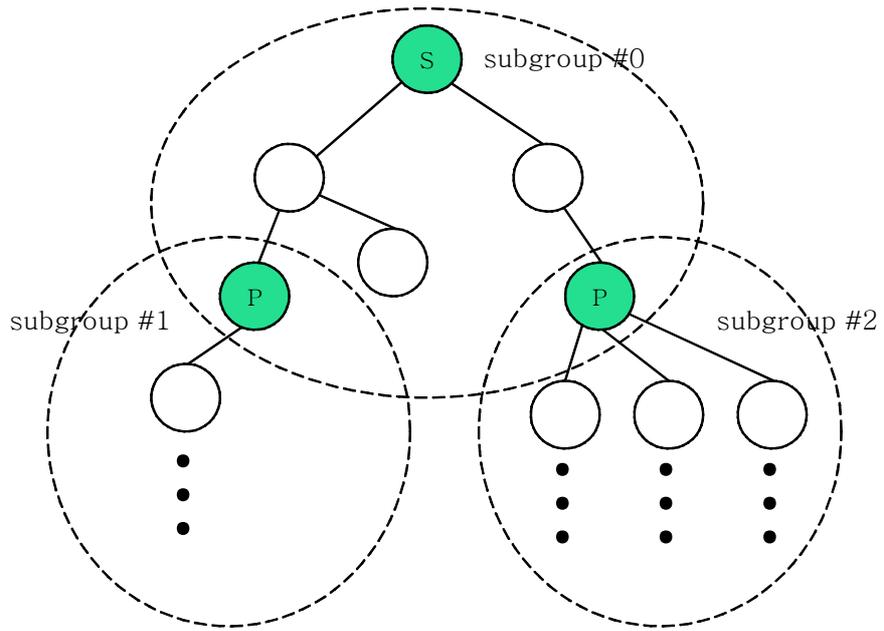

**Fig. 1** HRM Model

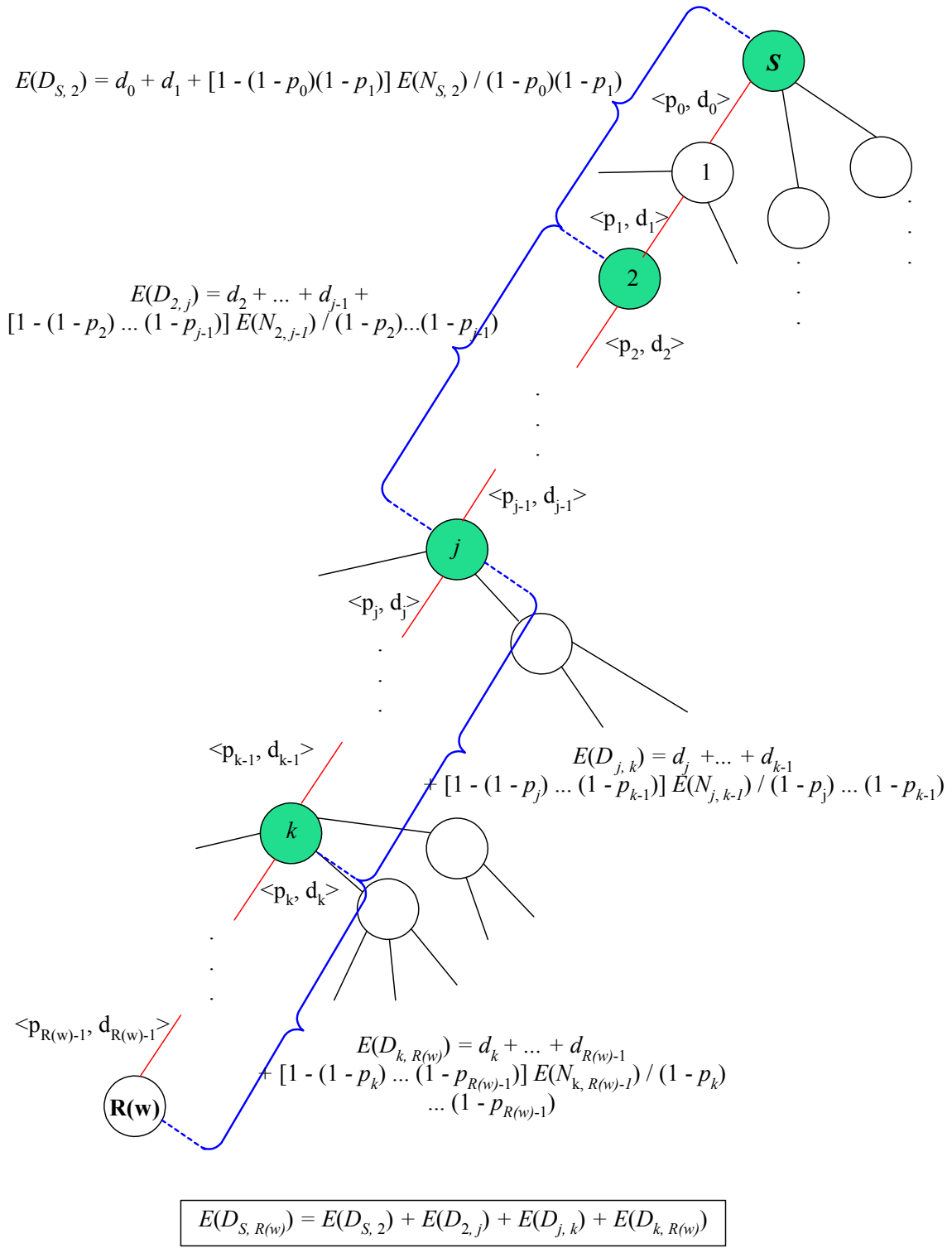

**Fig. 2** Expected delivery delay model reflecting heterogeneity: Expected delivery delay between node *S* and *R(w)* when at each node *S*, 2, *j* and *k*, a repair server is located.

***HMaxDelay***(*T(V, E)*, repair_server_size *k-1*) {

1. $V_L$=all leaf nodes in *T*;
2. list ***L***=*V*-$V_L$;
3. list ***Repair_Server_Set***=ø;
4. ***Compute_Expected_Delay***(*e* ∈*E*) /*Compute the expected delivery delay of each edge *e* ∈*E**/
5. *server_size* = 0;
6. while (*server_size* < *k-1* || ***L*** is not empty) {
7.    *list_index* = ***Find_Largest_Delay***(***L***); /*the index in ***L*** whose outbound incident edge has the largest expected delivery delay*/
8.    if (***L***[*list_index*] is not the root) {
9.      ***Repair_Server_Set***[*server_size*] = ***L***[*list_index*];
10.      Clear ***L***[*list_index*];
11.      *server_size*++;
12.    } else
13.      Clear ***L***[*list_index*];
   }
}

**Fig. 3** Heuristic Algorithm of *HMaxDelay*

**HMaxDegree**(*T(V, E)*, repair_server_size *k-1*) {

1. $V_L$=all leaf nodes in *T*;
2. list ***L***=*V*-$V_L$;
3. list ***Repair_Server_Set***=ø;
4. *server_size* = 0;
5. while (*server_size* < *k-1* || ***L*** is not empty) {
6.    *list_index* = **Find_Largest_Degree**(***L***); /*the index in ***L*** who has the highest degree*/
7.    if (***L***[*list_index*] is not the root) {
8.      ***Repair_Server_Set***[*server_size*] = ***L***[*list_index*];
9.      Clear ***L***[*list_index*];
10.      *server_size*++;
11.    } else
12.      Clear ***L***[*list_index*];
   }
}

**Fig. 4** Heuristic Algorithm of *HMaxDegree*

**HLongPath**(*T(V, E)*, repair_server_size *k-1*) {
1. $V_L$ = all leaf nodes in *T*;
2. ***P*** = a set of paths between the root and all $v \in V_L$;
3. list ***Repair_Server_Set***=ø;
4. *server_size* = 0;
5. while (*server_size* < *k-1*) || (***L*** is not empty) {
6.     ***Find_Largest_Path***($p_i \in$***P***); /*Find the path $p_i \in$ ***P*** that has the largest expected delivery de-
7. lay*/
8.     ***Locate_Repair_Server***($p_i \in$ ***P***); /*Assume that repair servers are located in all nodes but $v \in V_L$
9. on the path $p_i$*/
10.     ***Compute_Expected_Delay***($e \in p_i$); /*Compute expected delivery delay of each edge on the path $p_i$*/
11.     $V_y$ = all nodes on the path $p_i$;
12.     $v_j$=***Find_Unmarked***($V_y$); /*Find a unmarked node $v_i$ in $V_y$ such that its outbound incident edge has the largest expected delivery delay*/
13.     if ($v_j$ found) {
14.         ***Repair_Server_Set***[*server_size*] = $v_i$;
15.         Mark $v_i$;
16.         *server_size*++;
        }
    }
}

**Fig. 5** Heuristic Algorithm of *HLongPath*

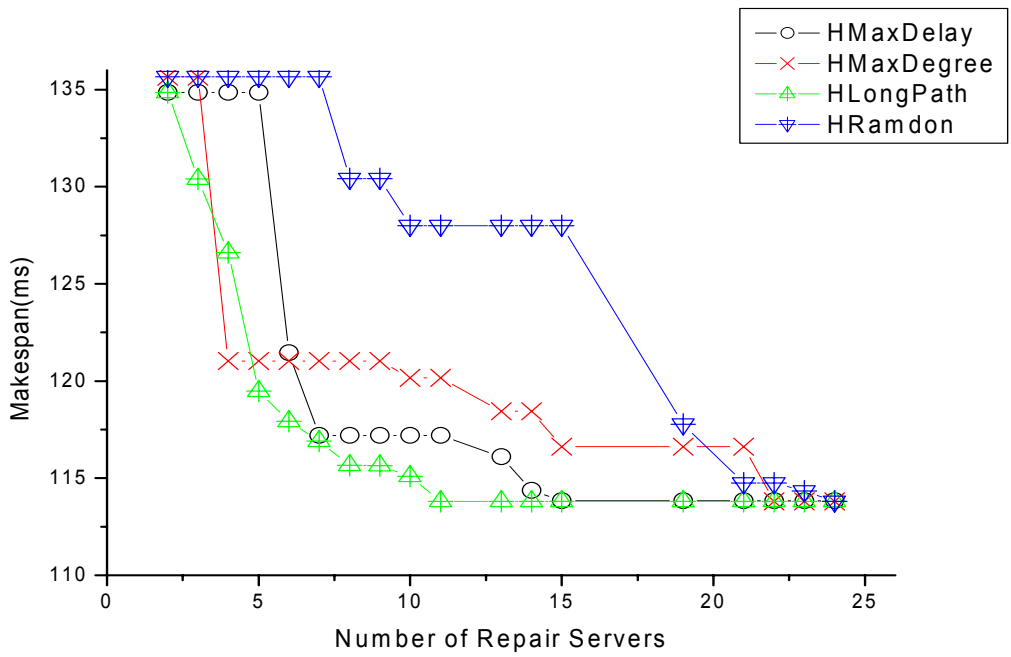

(a) In case network size of 100 nodes

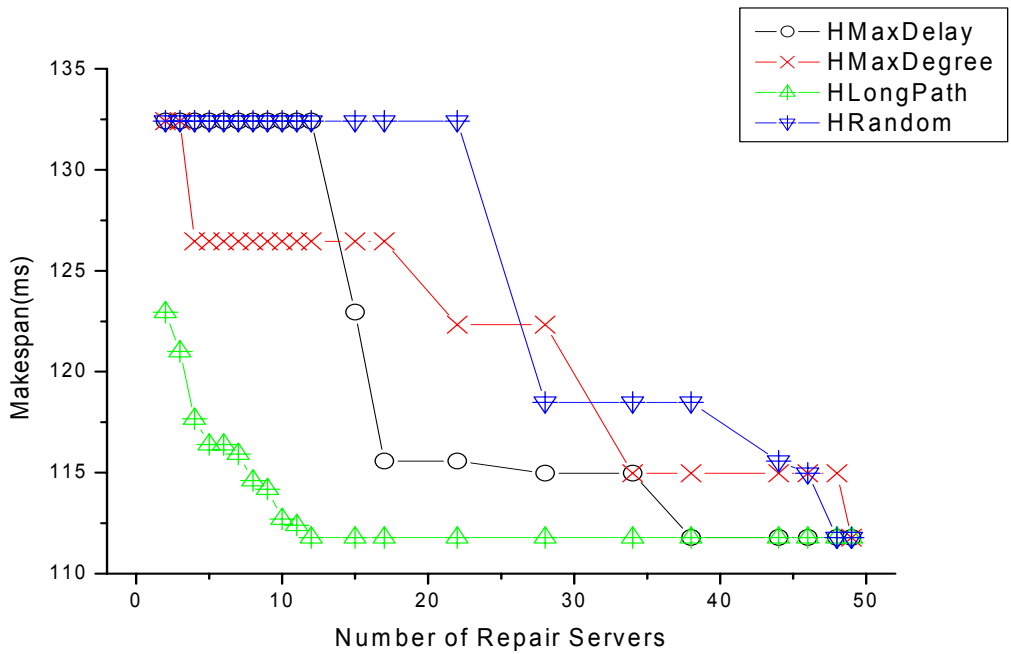

(b) In case network size of 200 nodes

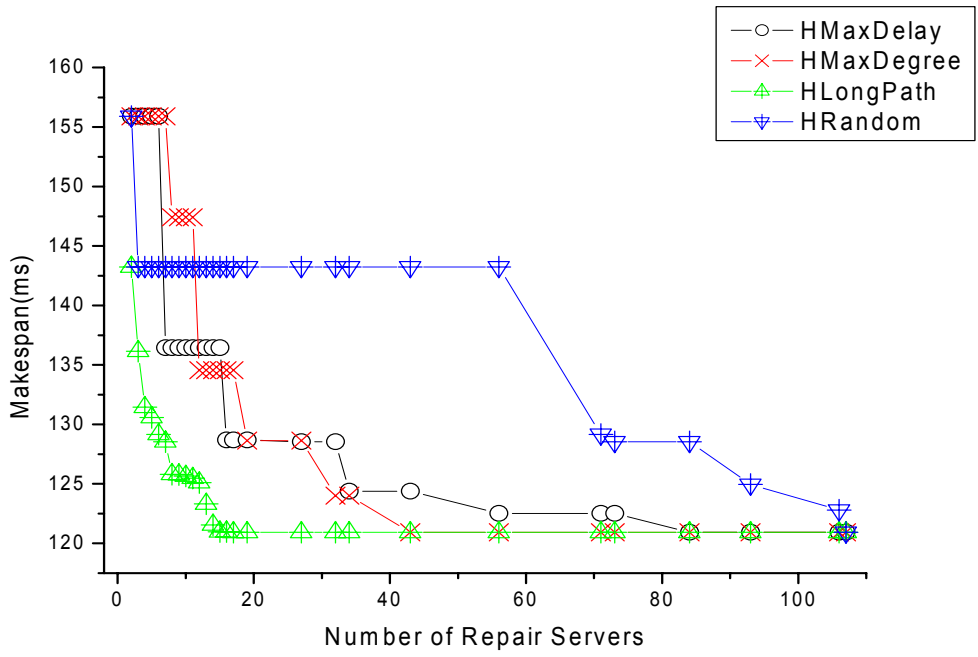

(c) In case network size of 500 nodes

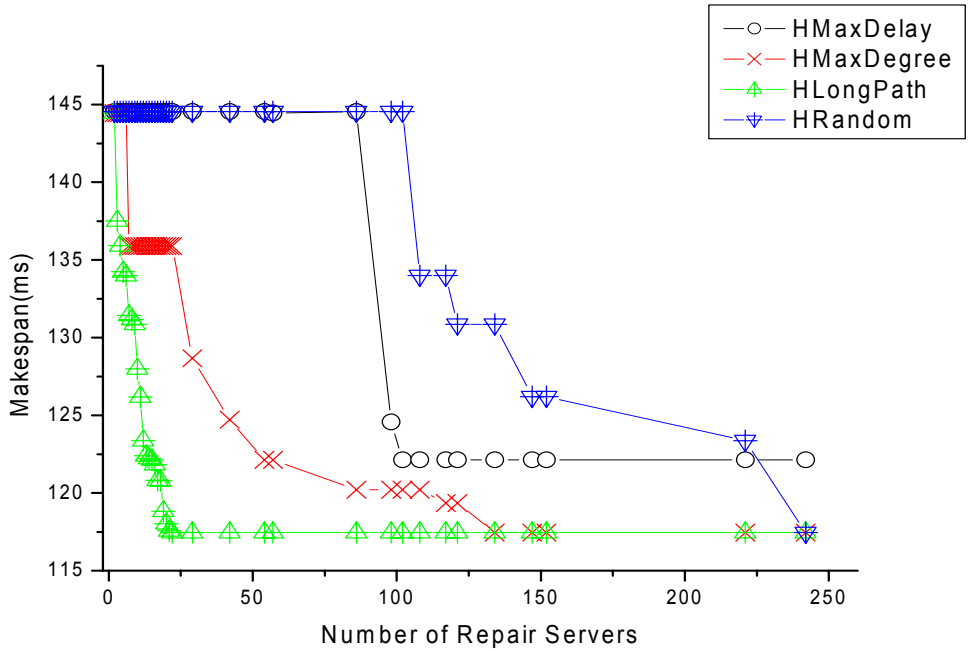

(d) In case network size of 1000 nodes

**Fig. 6** Makespan of each heuristic as a function of the number of repair servers

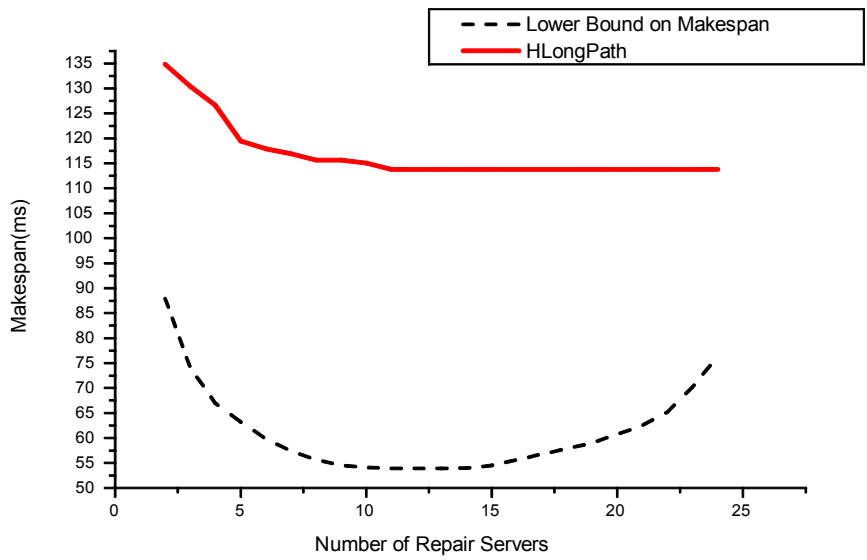

**(a) Comparison with lower bound on makespan**

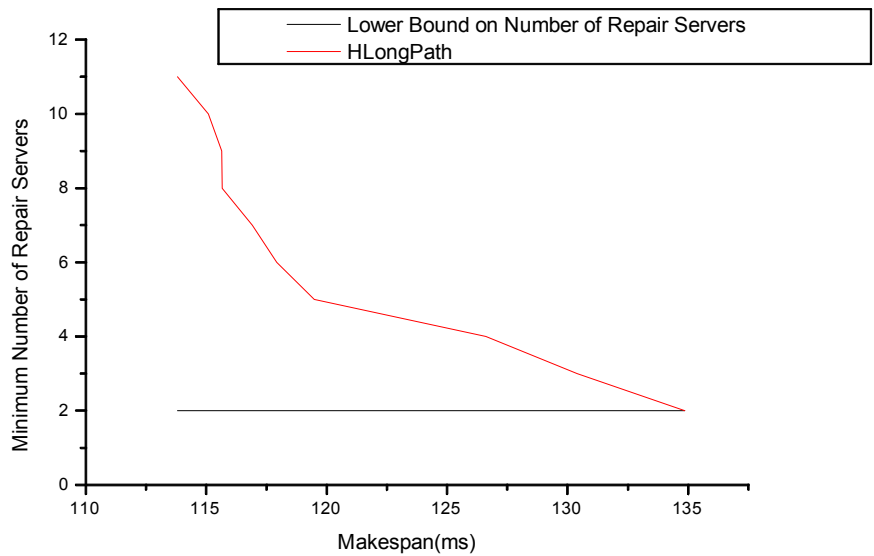

**(b) Comparison with lower bound on number of repair servers**

**Fig. 7** Comparison of *HLongPath* in network size of 100 nodes

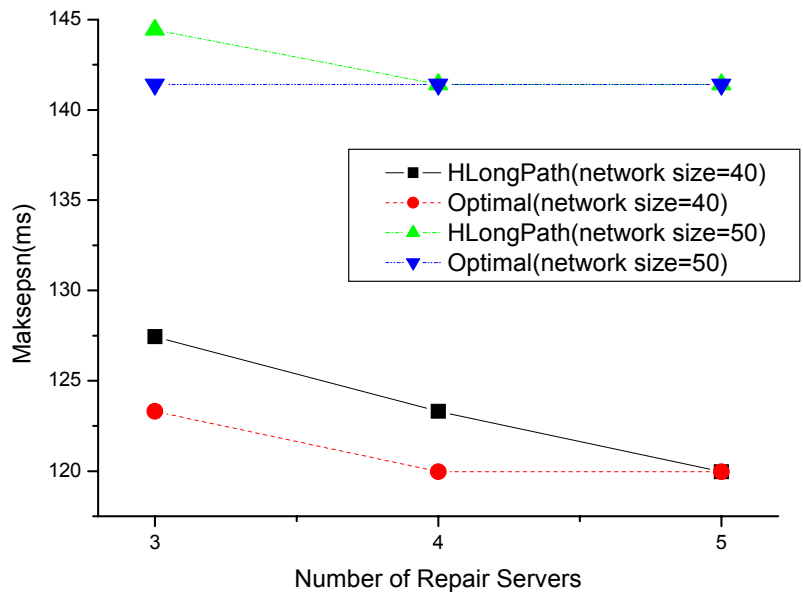

**Fig. 8** Comparison with optimal in network size of 40 and 50nodes

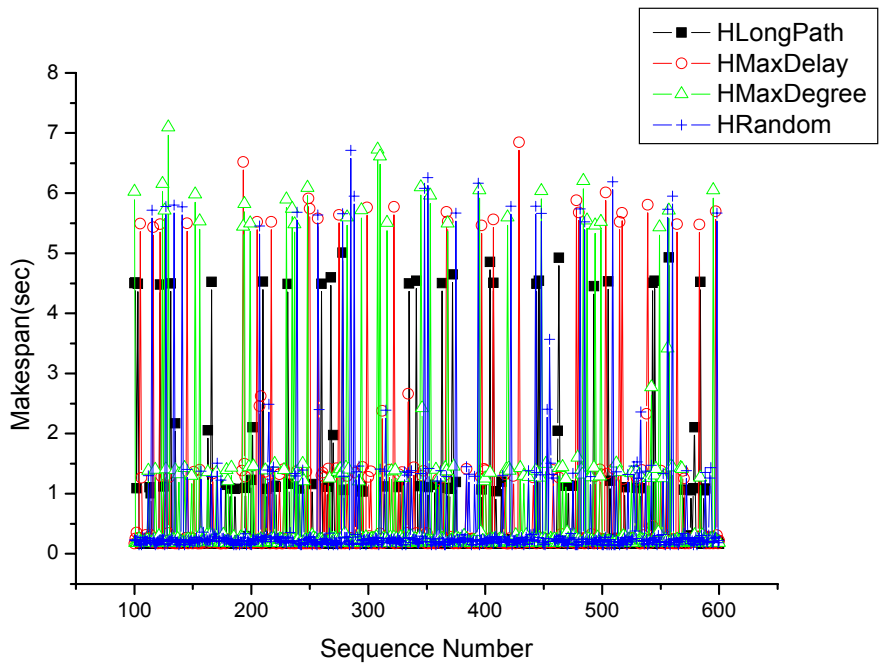

**Fig. 9** Makespan of each heuristics measured on PGM

**Table 1**. Notations for expected delivery model

| Value | Description |
|---|---|
| $S$ | Sender node |
| $R(w)$ | Receiver node |
| $L_{a,b}$ | Summation of per-link delays on all links between node *a* and node *b* |
| $E(N_{a,b})$ | When node *a* acts as a repair server of node *b*, the expected inter-arrival time of two consecutive packets in node *b*. (their sequence numbers are consecutive) |
| $\pi(a,b)$ | Set of all repair servers located on the path between node *a* and node *b*. |
| $\theta_{a,b}$ | Number of elements in set $\pi(a,b)$ |
| $E(D_{S,R(w)})$ | Expected delivery delay between sender node *s* and receiver node *R(w)* |
| $\psi(a,b)$ | Set of all links between node *a* and node *b*. |

**Table 2**. Each tuple - (**Lower limit of makespan (*ms*), minimum number of repair servers to attain this lower limit**) in accordance with each heuristic and network size.

| Heuristic | Network Size | | | |
|---|---|---|---|---|
| | 100 (25) | 200 (50) | 500 (113) | 1000 (244) |
| *HRandom* | (113.81, 18) | (111.78, 48) | (120.92, 107) | (117.46, 242) |
| *HMaxDelay* | (113.84, 13) | (111.78, 38) | (120.92, 84) | (122.14, 102) |
| *HMaxDegree* | (113.81, 22) | (111.78, 49) | (120.92, 43) | (117.46, 134) |
| *HLongPath* | (113.81, 11) | (111.78, 12) | (120.92, 16) | (117.46, 22) |

**Table 3.** Makespan of our heuristic and the integer optimal solution

| Proxy Size | | Network Size | | | |
|---|---|---|---|---|---|
| | | 25 | 30 | 40 | 50 |
| 3 | optimal | 123.31 | 123.31 | **123.31** | **141.41** |
| | greedy | 123.31 | 123.31 | **127.44** | **144.43** |
| 4 | optimal | 119.96 | 119.96 | **119.96** | 141.41 |
| | greedy | 119.96 | 119.96 | **123.31** | 141.41 |
| 5 | optimal | - | - | 119.96 | - |
| | greedy | - | - | 119.96 | - |

**Table 4** Simulation parameters

| Parameter | Value |
| --- | --- |
| Topology size | 200 nodes |
| Number of receiver | 44 |
| Number of proxy to be located | 10 |
| Topology generator | ToGend |
| Per-packet size | 1000 byte |
| Traffic load | 500 kb/s |
| Simulation duration | 30 seconds |
| Per-link loss rate | Varies 0.1 ~ 0.001 |
| Per-link propagation delay | Varies 10ms ~ 40ms |